\documentstyle[preprint,aps,prb]{revtex}
\def\be{\begin{equation}}
\def\ee{\end{equation}}
\begin{document}
\draft 
\title{The Role of the Korteweg--de Vries Hierarchy in the N--Soliton 
Dynamics of the Shallow Water Wave Equation}
\author{R. A. Kraenkel$^1$, M. A. Manna$^2$, J. C. Montero$^1$ and J. G. 
Pereira$^1$ }
\vskip 0.5cm
\address{$^1$Instituto de F\'{\i}sica Te\'orica\\
Universidade Estadual Paulista\\
Rua Pamplona 145\\
01405-900\, S\~ao Paulo \\ 
Brazil
\vskip 1.0cm 
$^2$Physique Math\'ematique et Th\'eorique, UPRES--A 5032\\
Universit\'e de Montpellier II\\
34095 Montpellier Cedex 05\\
France}
\maketitle
\begin{abstract}
We apply a multiple--time version of the reductive perturbation method
to study long waves as governed by the shallow water wave model 
equation. As a consequence of the requirement of a secularity--free 
perturbation theory, we show that the well known N--soliton dynamics of 
the shallow water wave equation, in the  particular case of $\alpha=2 \beta$, 
can be reduced to the N--soliton solution that satisfies simultaneously all 
equations of the Korteweg--de Vries hierarchy.
\end{abstract}
\vspace{1.0 cm}
{\hspace{1.7 cm} Keywords: Perturbation Theory, Secularity, KdV Hierarchy}
\vfill \eject

\section{Introduction}

The shallow water wave (SWW) equation,
\be
u_{xxxt} + \alpha u_x u_{xt} + \beta u_t u_{xx} - u_{xt} - u_{xx} = 0 \, ,
\label{sww}
\ee
is known to present solitonic solutions for two families of values of the
parameters $\alpha$ and $\beta$. The first family, given by
\be
\alpha = 2 \beta \, ,
\label{a2b}
\ee
was shown to be integrable by the inverse scattering method by Ablowitz, 
Kaup, Newell and Segur.$^{1)}$ The second family, given by
\be
\alpha = \beta \, ,
\label{ab}
\ee
was shown to possess N soliton solutions by Hirota and Satsuma.$^{2)}$ In
addition, it has been shown$^{3,4)}$ that Eq.(\ref{sww}) passes the Painlev\'e 
test if and only if either condition (\ref{a2b}) or (\ref{ab}) is satisfied, which is 
an indication that this equation might be integrable only for these two cases. 
We would like also to remark that Eq.(\ref{sww}) may be written as an 
integrodifferential equation:
\be
U_{xxt} + \alpha U U_t - \beta U_x \int_{x}^{\infty} U_t dx' - U_t - U_x = 
0 \, .
\label{isww}
\ee
In this form, the SWW equation was shown to be derivable in the Boussinesq 
approximation from the classical shallow water theory,$^{4)}$ but with the 
condition
\be
\beta = 2 \alpha \, ,
\ee
which is neither of the above discussed cases.

In this paper, we are going to apply a multiple--time$^{5,6)}$ reductive 
perturbation method$^{7)}$ to study the long--wave limit of the SWW 
equation. This method, when applied to nonlinear weak dispersive general 
systems, introduces a connection between the equations of the 
Korteweg--de Vries (KdV) hierarchy and the elimination of the 
soliton--related secularities appearing in the evolution equations of the 
higher--order terms of the wave--field. As a consequence of this approach, 
we will show that the well known solitary--wave solution to Eq.(\ref{sww}), 
existing for any $\alpha$ and $\beta$, can be written as a solitary--wave 
satisfying simultaneously all equations of the KdV hierarchy. This result is a 
consequence of an intrinsic property of the KdV solitary--wave, which allows 
for a truncation of the perturbative series. Then, we will show that, for the 
case $\alpha=2 \beta$, the N soliton dynamics of the SWW equation can be 
reduced to the dynamics of the N soliton solution of the KdV hierarchy. This 
result$^{8)}$ comes from the fact that the SWW equation with $\alpha=2 
\beta$, and the KdV equation are both related to the same eigenvalue 
problem in the AKNS scheme.$^{1)}$ 
On the other hand, for $\alpha=\beta$, the dynamics underlying the N soliton 
solutions of the SWW equation has a different nature, being similar to the 
dynamics governing the N soliton solution of the Boussinesq 
equation.$^{3)}$ This case will not be considered in this paper, which is 
organized as follows. In Section II we introduce the multiple--time 
perturbation scheme based on the long--wave expansion of the dispersion 
relation. In Section III we obtain the first few evolution equations, and in 
Section IV we show how the KdV hierarchy equations appear as a 
compatibility condition. The case of a solitary--wave solution for the KdV 
hierarchy is considered in Section V, and in Section VI we describe how the N 
soliton solution of the SWW equation with $\alpha=2 \beta$ can be obtained 
as the N soliton solution of the KdV hierarchy. And finally, in 
Section VII we summarize and discuss the results obtained.

\section{Long--Wave Expansion and the Multiple Times}

The linear dispersion relation of the SWW equation is given by
\be
\omega(k) = \frac{k}{1 + k^2} \, .
\label{dr}
\ee
To study its long--wave limit, we put
\be
k = \epsilon \kappa \, ,
\label{lw}
\ee
with $\epsilon$ a small parameter, and we expand the dispersion relation 
according to
\be
\omega(\kappa) = \epsilon \kappa - \epsilon^3 \kappa^3 + \epsilon^5 
\kappa^5
- \epsilon^7 \kappa^7 + \cdots \, .
\label{exp}
\ee
Introducing this expansion into the plane wave solution of the linear SWW 
equation, we get
\be
u = a \; {\rm exp} \, i \left[ \kappa \epsilon (x - t) + \kappa^3 \epsilon^3 t
- \kappa^5 \epsilon^5 t + \kappa^7 \epsilon^7 t - \cdots \right] \, ,
\label{ls2}
\ee
where $a$ is a constant. Inspired by this solution, we introduce a slow space
\be
\xi = \epsilon (x - t) \, ,
\label{ss}
\ee
as well as an infinite sequence of properly normalized slow time variables
\be
\tau_3 = \epsilon^3 t \quad; \quad \tau_5 = - \epsilon^5 t \quad; \quad
\tau_7 = \epsilon^7 t \quad; \quad {\rm etc}.
\label{st}
\ee
Accordingly, we have that
\be
\frac{\partial}{\partial x} = \epsilon \frac{\partial}{\partial \xi}
\label{dx}
\ee
and
\be
\frac{\partial}{\partial t} = - \epsilon \frac{\partial}{\partial \xi} +
\epsilon^3 \frac{\partial}{\partial \tau_3} -
\epsilon^5 \frac{\partial}{\partial \tau_5} +
\epsilon^7 \frac{\partial}{\partial \tau_7} - \cdots \, .
\label{dt}
\ee
As we are going to see later, the slow time normalizations introduced in the 
definitions (\ref{st}) are crucial in the sense that they will allow for an 
automatic elimination of the secularities appearing in the evolution equations 
of the higher order terms of the wave--field.$^{9)}$

\section{Multiple Time Evolution Equations}

We expand now the wave--field $u$ according to
\be
u = \epsilon {\hat u} = \epsilon \left(u_0 + \epsilon u_1 + \epsilon^2 u_2
+ \cdots \right) \, ,
\label{uhat}
\ee
and substitute it, together with Eqs.(\ref{dx}) and (\ref{dt}), in the SWW
equation (\ref{sww}). The result is:
\begin{eqnarray}
\frac{\partial^3}{\partial \xi^3} \Big(- \epsilon 
\frac{\partial}{\partial \xi} &+&
\epsilon^3 \frac{\partial}{\partial \tau_3} -
\epsilon^5 \frac{\partial}{\partial \tau_5} + \cdots \Big) {\hat u} -
\frac{\partial}{\partial \xi} 
\Big(\epsilon \frac{\partial}{\partial \tau_3} -
\epsilon^3 \frac{\partial}{\partial \tau_5} + \cdots \Big) {\hat u}
\nonumber \\
&+& \alpha \frac{\partial {\hat u}}{\partial \xi}
\Big(- \epsilon \frac{\partial^2}{\partial \xi^2} +
\epsilon^3 \frac{\partial}{\partial \xi} \frac{\partial}{\partial \tau_3} -
\epsilon^5 \frac{\partial}{\partial \xi} \frac{\partial}{\partial \tau_5} + \cdots 
\Big) 
{\hat u} \nonumber \\ 
&+&\beta \frac{\partial^2 {\hat u}}{\partial \xi^2} \Big(- \epsilon 
\frac{\partial}{\partial \xi} +
\epsilon^3 \frac{\partial}{\partial \tau_3} -
\epsilon^5 \frac{\partial}{\partial \tau_5} + \cdots \Big) {\hat u} = 0 \, .
\label{esww}
\end{eqnarray}  

We proceed then to an order--by--order inspection of this equation. At 
order $\epsilon^0$ we get
\be
\frac{\partial^4 u_0}{\partial \xi^4} + \frac{\partial^2 u_0}{\partial \xi
\partial \tau_3} + (\alpha + \beta) \frac{\partial u_0}{\partial \xi}
\frac{\partial^2 u_0}{\partial \xi^2} = 0 \, .
\label{zero}
\ee
We make now, in all components $u_n$ of the wave--field expansion, the 
following transformation:
\be
\frac{\partial u_n}{\partial \xi} = - \frac{6}{\alpha + \beta} \,\, v_n \quad ; \quad  
n=0, 1, 2, \dots \; \; .
\label{trans}
\ee
Consequently, Eq.(\ref{zero}) acquires the form
\be
\frac{\partial v_0}{\partial \tau_3} + \frac{\partial^3 v_0}{\partial \xi^3} -
6 v_0 \frac{\partial v_0}{\partial \xi} = 0 \, ,
\label{kdv}
\ee
which is the KdV equation in the time $\tau_3$.

At order $\epsilon^1$, Eq.(\ref{esww}) gives:
\be
\frac{\partial^4 u_1}{\partial \xi^4} + \frac{\partial^2 u_1}{\partial \xi \partial 
\tau_3} + 
(\alpha + \beta) \frac{\partial u_0}{\partial \xi} \frac{\partial^2 u_1}{\partial 
\xi^2} + 
(\alpha + \beta) \frac{\partial u_1}{\partial \xi} \frac{\partial^2 u_0}{\partial 
\xi^2} = 0 \, .
\label{um}
\ee
Transforming $u_0$ and $u_1$ according to Eq.(\ref{trans}), we obtain
\be
\frac{\partial v_1}{\partial \tau_3} + \frac{\partial^3 v_1}{\partial \xi^3} -
6 \frac{\partial}{\partial \xi} (v_0 v_1) = 0 \, ,
\label{kdv1}
\ee
that is, $v_1$ satisfies a homogeneous linearized KdV equation in the time 
$\tau_3$.

At order $\epsilon^2$, Eq.(\ref{esww}) gives:
\begin{eqnarray}
- \frac{\partial^4 u_2}{\partial \xi^4} + \frac{\partial^4 u_0}{\partial \xi^3 \partial 
\tau_3} - 
\frac{\partial^2 u_2}{\partial \xi \partial \tau_3} +
\frac{\partial^2 u_0}{\partial \xi \partial \tau_5} &{}& \nonumber \\ 
-
(\alpha + \beta) \left(\frac{\partial u_0}{\partial \xi} \frac{\partial^2 u_2}{\partial 
\xi^2} + 
\frac{\partial u_1}{\partial \xi} \frac{\partial^2 u_1}{\partial \xi^2} + \frac{\partial 
u_2}
{\partial \xi} \frac{\partial^2 u_0}{\partial \xi^2} \right) &+& \alpha \frac{\partial 
u_0}
{ \partial \xi}
\frac{\partial^2 u_0}{\partial \xi \partial \tau_3} + \beta \frac{\partial^2 
u_0}{\partial \xi^2} 
\frac{\partial u_0}{\partial \tau_3} = 0 \, .
\label{dois}
\end{eqnarray}
Transforming $u_0$, $u_1$ and $u_2$ according to Eq.(\ref{trans}), and using 
the KdV equation 
(\ref{kdv}) to express $v_{0\tau_3}$, we obtain
\begin{eqnarray}
\frac{\partial v_2}{\partial \tau_3} &+& \frac{\partial^3 v_2}{\partial \xi^3} -
6 \frac{\partial}{\partial \xi} (v_0 v_2) = \frac{\partial v_0}{\partial \tau_5} + 
\left[\frac{18 \alpha + 24 \beta}{\alpha + \beta}\right] \frac{\partial v_0}{\partial 
\xi} 
\frac{\partial^2 v_0}{\partial \xi^2} \nonumber \\ 
&+& \left[\frac{12 \alpha + 6 \beta}{\alpha + \beta}\right] v_0 \frac{\partial^3 
v_0}{\partial \xi^3} - 
\left[\frac{36 \alpha + 18 \beta}{\alpha + \beta}\right] {v_0}^2 \frac{\partial 
v_0}{\partial \xi} - 
\frac{\partial^5 v_0}{\partial \xi^5} + 6 v_1 \frac{\partial v_1}{\partial \xi} \; . 
\label{v2}
\end{eqnarray}
This equation involves the evolution of $v_0$ in the time $\tau_5$, which is 
not known up to this point. Moreover, when $v_0$ is assumed to be a 
solitary--wave solution of the KdV equation, the source term ${\partial^5 
v_0}/{\partial \xi^5}$ will be a secular producing term.  Therefore, 
before continuing we have to solve these two problems.

\section{The KdV Hierarchy as a Compatibility Condition}

As we have already seen, $v_0$ satisfies the KdV equation in the time 
$\tau_3$. The evolution of $v_0$ in the higher order times $\tau_{2n+1}$ can 
then be obtained in the following way.$^{5)}$ First, to have a well ordered 
perturbative scheme, we impose that each one of these equations be 
$\epsilon$--independent when passing from the slow $(v_0, \xi, 
\tau_{2n+1})$ to the laboratory coordinates $(v, x, t)$. This will select all 
possible terms to appear in the equation for $v_{0\tau_{2n+1}}$.  Then, by 
imposing the natural (in the multiple time formalism) compatibility 
condition
\be
\Big(v_{0\tau_3} \Big)_{\tau_{2n+1}} = \Big(v_{0\tau_{2n+1}} \Big)_{\tau_3} \, ,
\label{compa}
\ee
it is possible to determine the above constants in terms of $\alpha_{2n+1}$, 
which is left as a free-parameter. As it can be easily verified through an 
explicit calculation,$^{5)}$ the resulting equations are exactly those given 
by the KdV hierarchy. As an example, let us consider the evolution of $v_0$ 
in time $\tau_5$. From the $\epsilon$-independence requirement, 
$v_{0\tau_5}$ is restricted to be of the form
\be
v_{0\tau_5} = \alpha_5 v_{0(5\xi )} + \beta_5 v_0 v_{0\xi \xi \xi } + \gamma_5 
v_{0\xi }v_{0\xi \xi } + \delta_5 v_0^2 v_{0\xi } \, ,
\ee
where $\alpha_5$, $\beta_5$, $\gamma_5$ and $\delta_5$ are constants. The 
compatibility condition 
$$
\Big(v_{0\tau_3} \Big)_{\tau_{5}} = \Big(v_{0\tau_{5}} \Big)_{\tau_3} \, ,
$$
then, determines $\beta_5$, $\gamma_5$ and $\delta_5$ in terms of 
$\alpha_5$, which is left as a
free-parameter, yielding the equation
\be
v_{0\tau_5} = v_{0(5\xi)} - 10 v_0 v_{0\xi\xi\xi} - 20 v_{0\xi} v_{0\xi\xi} +
30 {v_0}^2 v_{0\xi} \, .
\label{kdv5}
\ee
Analogously, for the evolution of $v_0$ in $\tau_7$, we obtain
\begin{eqnarray}
v_{0\tau_7} = &-& v_{0(7\xi)} + 14 v_0 v_{0(5\xi)} + 42 v_{0\xi} v_{0(4\xi)} + 140 
(v_0)^3 v_{0\xi} 
\nonumber \\ 
&+& 70 v_{0\xi\xi} v_{0\xi\xi\xi} - 280 v_0 v_{0\xi} v_{0\xi\xi} - 70(v_{0\xi})^3 - 
70 {v_0}^2 v_{0\xi\xi\xi} \, .
\label{kdv7}
\end{eqnarray}
Equations (\ref{kdv5}) and (\ref{kdv7}) are the first two higher--order 
equations of the KdV hierarchy. The same procedure can be used to 
generate any higher-order equation of the KdV hierarchy. The right--hand 
side of these equations would in principle appear multiplied by the 
corresponding free--parameter left at each order, which would account for 
different possible slow time normalizations. However, since we have already 
defined appropriate slow time normalizations, these parameters were taken 
to be $1$ in order to have an agreement with the normalizations introduced in 
Eq.(\ref{st}). This is an important point since it will allow for an automatic 
elimination$^{9)}$ of the soliton related secular producing terms present in 
the higher-order perturbation equations, which are always 
of the form$^{10)}$
\be
v_{0(2n+1)\xi} \quad; \quad n = 0,1,2, \dots \, \, .
\ee

\section{The Solitary--Wave Solution}

In this section, we are going to consider some specific solutions to our 
equations. First, we will assume for $v_1$ the trivial solution
\be
v_1 = 0 \, .
\ee
Then, by taking into account that the evolution of $v_0$ in the time $\tau_5$ 
is now known to proceed according to the fifth--order equation of the KdV 
hierarchy, given by Eq.(\ref{kdv5}), we obtain for Eq.(\ref{v2})
\be
v_{2\tau_3} + v_{2\xi\xi\xi} - 6 (v_0 v_2)_{\xi} = \frac{2 \alpha - 4 \beta}
{\alpha + \beta} \left[ - 3 {v_0}^2 v_{0\xi} + v_0 v_{0\xi\xi\xi} - v_{0\xi}
v_{0\xi\xi} \right] \, .
\label{v2b}
\ee
We see in this way that the use of the properly normalized KdV hierarchy 
equation to express $v_{0\tau_5}$ has automatically canceled out the secular 
producing term $v_{0(5\xi)}$ from Eq.(\ref{v2}). Therefore, besides satisfying 
the KdV equation in the time $\tau_3$, the wave--field $v_0$ must satisfy 
also the fifth--order KdV hierarchy equation (\ref{kdv5}) in the time $\tau_5$. 
Actually, in order to have a secularity--free perturbation theory up to any 
higher order, $v_0$ will be required to satisfy simultaneously all equations 
of the KdV hierarchy, each one in a different slow time coordinate. 

Next, we assume for $v_0$ a KdV solitary--wave solution. As $v_0$
must satisfy simultaneously all equations of the KdV hierarchy, we take
\be
v_0 = - 2 \kappa^2 {\rm sech^2} \left[ \kappa \xi - 4 \kappa^3 \tau_3 + 16 
\kappa^5 \tau_5 - 64 \kappa^7 
\tau_7 + \cdots \right] \, .
\label{eter}
\ee
Now, using this solitary--wave solution, one can easily check that the 
right--hand side of Eq.(\ref{v2b}) vanishes, implying that, along with $v_1$, 
$v_2$ also satisfies a homogeneous linearized KdV equation:
\be
v_{2\tau_3} + v_{2\xi\xi\xi} - 6 (v_0 v_2)_{\xi} = 0 \, .
\ee
Like in the case of $v_1$, we will assume for it the trivial solution
\be
v_2 = 0 \, .
\ee

We proceed now to the next order, in which case Eq.(\ref{esww}) gives
\be
u_{3(4\xi)} - u_{3 \xi \tau_3} + (\alpha + \beta) \left(u_{0\xi} u_{3\xi} \right)_{\xi} 
= 0 \, ,
\ee
where, due to the fact that $v_1 = v_2 = 0$, we have already used that 
$u_1 = u_2 = 0$. Then, by transforming $u_0$ and $u_3$ according to 
Eq.(\ref{trans}), we get the following equation for $v_3$:
\be
v_{3 \tau_3} + v_{3 \xi \xi \xi} - 6 \left(v_{0} v_{3} \right)_{\xi} = 0 \, .
\ee
We see in this way that also $v_3$ satisfies a homogeneous linearized KdV 
equation, and as in the previous cases we choose for it the trivial solution
\be
v_3 = 0 \, .
\ee

This is actually what will happen at every higher order: after using the 
equations of the KdV hierarchy to express $v_{0\tau_{2n+1}}$, and after 
assuming the solitary--wave solution (\ref{eter}) 
for $v_0$, the evolution of $v_n$ in the time $\tau_3$ will be given by a 
homogeneous linearized KdV equation. Consequently, it will always be 
possible to choose the trivial solution
\be
v_n = 0 \quad ; \quad n = 1, 2, 3, \dots \, \, .
\label{000}
\ee
It should be mentioned that the evolution of $v_n$ $(n=1,2,3,\dots)$ in each 
higher--order time $\tau_{2n+3}$ is always given by a {\it linear} partial 
differential equation, which becomes homogeneous after the choice 
$v_{n-1}=0$. Therefore, the solution (\ref{000}) is in indeed always possible 
and compatible with the whole perturbative scheme. 

We return now to the laboratory coordinates $(x,t,u)$. First, we recall that we 
have assumed for $u$ the expansion given by Eq.(\ref{uhat}).
Moreover, we have obtained a particular solution in which
$u_1 = u_2 = u_3 = \cdots = 0 $.
Therefore, the laboratory wave--field $u$ is given simply by
\be
u = \epsilon u_0 \, .
\label{wf}
\ee 
On the other hand, $v_0$ was taken to be given by Eq.(\ref{eter}), which is a 
solitary--wave satisfying all equations of the KdV hierarchy. By using the 
transformation (\ref{trans}), we can obtain the corresponding $u_0$:
\be
u_0 = - \frac{12 \kappa^2}{\alpha + \beta} \int_{\xi}^{\infty} {\rm sech}^2
\left[ \kappa \xi' - 4 \kappa^3 \tau_3 + 16 \kappa^5 \tau_5 - 64 \kappa^7 
\tau_7
+ \cdots \right] d{\xi}' \, .
\ee
Integrating and using Eq.(\ref{wf}), we get
\be
u = - \frac{12 \epsilon \kappa}{\alpha + \beta}\left[1 - {\rm tanh}(\kappa \xi -
4 \kappa^3 \tau_3 + 16 \kappa^5 \tau_5 - 64 \kappa^7 \tau_7 + 
\cdots) \right] \, .
\ee
We drive now $u$ back to the laboratory by eliminating the slow variables 
$(\kappa, \xi, \tau_{2n+1})$, which are related to the laboratory ones
$(k, x, t)$ according to equations (\ref{lw}), (\ref{ss}) and (\ref{st}).
The result is
\be
u = - \frac{12 k}{\alpha + \beta}\left[1 - {\rm tanh}\left(k x -
k\{1 + 4 k^2 + 16 k^4 + 64 k^6 + \cdots \} t \right) \right] \, .
\ee
But, the series appearing inside the curly brackets can be summed, yielding 
finally
\be
u = - \frac{12 k}{\alpha + \beta}\left[1 - {\rm tanh} \left(
k x - \frac{k t}{1 - 4 k^2} \right) \right] \, ,
\ee
which is the well known solitary--wave solution of the SWW equation 
(\ref{sww}), valid for any value of $\alpha$ and $\beta$, except of course 
$\alpha=-\beta$ .

\section{Two--or--More Soliton Solutions}

The elimination of secular producing terms through the multiple--scale 
method is not always possible in the case where a N--soliton solution is 
assumed for the KdV equation because of the possible appearance of 
obstacles in the perturbative series.$^{11)}$
Obstacles are higher--order effects coming from the nonintegrability of the 
original equation which, when present, preclude the existence of uniform 
asymptotic expansions. An example of such a manifestation can be found in 
Ref.[12]. Here, however, we consider a particular case where no obstacles are 
present, namely the SWW equation with $\alpha=2 \beta$.

Let us then return to the perturbation theory, but considering now , 
the case in which $v_0$ is assumed to be the two--soliton solution of the 
KdV equation (\ref{kdv}), which is given by
\be
v_0 = - 4 (\kappa_{2}^{2} - \kappa_{1}^{2}) \left[ \frac{\kappa_{2}^{2} 
-\kappa_{1}^{2} + 
\kappa_{1}^{2} \cosh (2 \kappa_2 \gamma_2) + \kappa_{2}^{2} \cosh (2 
\kappa_1 \gamma_1)}
{\left[ (\kappa_2 - \kappa_1) \cosh (\kappa_1 \gamma_1 + \kappa_2 
\gamma_2) + (\kappa_2 + \kappa_1) 
\cosh (\kappa_1 \gamma_1 - \kappa_2 \gamma_2) \right]^2 } \right] \, ,
\label{2sol}
\ee
with
\be
\gamma_i = \xi - 4 \kappa_{i}^{2} \tau_3 \; \; ,  i = 1, 2 \, .
\label{gamai}
\ee
In this case, the whole perturbative scheme follows closely the one when 
$v_0$ is a one--soliton solution to the KdV equation. In particular, 
provided the evolution of $v_0$ in the time $\tau_5$ be given by the 
fifth--order KdV hierarchy equation (\ref{kdv5}), the equation for $v_2$ in the 
time $\tau_3$ will again be given by
\be
v_{2\tau_3} + v_{2\xi\xi\xi} - 6 (v_0 v_2)_{\xi} = \frac{2 \alpha - 4 \beta}
{\alpha + \beta} \left[ - 3 {v_0}^2 v_{0\xi} + v_0 v_{0\xi\xi\xi} - v_{0\xi}
v_{0\xi\xi} \right] \, .
\label{v2c}
\ee
However, when the two--soliton solution (\ref{2sol}) is substituted in its 
right--hand side, it does not vanish as it did when $v_0$ was assumed to be 
the solitary--wave solution (\ref{eter}). But, as can be easily seen, for the 
specific case of $\alpha= 2 \beta$, Eq.(\ref{v2c}) will be homogeneous 
independently of the solution chosen for $v_0$. Therefore, also for the case 
in which $v_0$ represents a two--soliton solution, it is possible to choose 
the trivial solution $v_2 = 0$. This is actually what happens at any higher 
order. In fact, to obtain a secularity--free perturbation theory, we first notice 
that $v_0$ must satisfy simultaneously all equations of the KdV hierarchy, 
which means that instead of (\ref{gamai}) we have ($ i = 1, 2$):
\be
\gamma_i = \xi - 4 \kappa_{i}^{2} \tau_3 + 16 \kappa_{i}^{4} \tau_5 - 64 
\kappa_{i}^{6} \tau_7 + \cdots \, .
\label{gamai2}
\ee
Then, for the case $\alpha=2 \beta$, the evolution of $v_n$ in the time 
$\tau_3$ will always be given by a homogeneous linearized KdV equation, 
whose solution we assume to be
$$
v_n = 0 \quad ; \quad n = 1, 2, 3, \dots \; .
$$
In other words, the perturbative series truncates, and an exact solution 
$u=\epsilon^2 u_0$ can be found also for a two--soliton solution.

The return to the laboratory coordinates $(x,t,u)$ proceeds in the very same 
way as for the case of the solitary--wave. Writing $(i = 1, 2)$
$$
\kappa_i = \epsilon^{-1} k_i \, ,
$$
and using Eqs.(\ref{ss}) and (\ref{st}), we find that, in the laboratory 
coordinates, the two soliton solution of the KdV hierarchy is given by
\be
u = \epsilon u_0 = - \frac{12}{\alpha + \beta} (k_1 + k_2) \left[ 1 + \frac{(k_1 - 
k_2) \tanh 
(k_2 \gamma_2)}{k_2 - k_1 \tanh (k_1 \gamma_1) \tanh (k_2 \gamma_2)} 
\right] \; ,
\label{2soll}
\ee
where now 
\be
\gamma_i = x - \frac{t}{1 - 4 k_{i}^{2}} \; \; \;   i = 1, 2 \, .
\ee
This is exactly the two--soliton solution of the SWW equation when 
$\alpha=2 \beta$. Therefore, like in the case of a solitary--wave, the 
two--soliton solution of the SWW equation is nothing but a 
two--soliton satisfying simultaneously, in the slow variables, all equations of 
the KdV hierarchy. Actually, as a similar procedure shows, this statement 
holds for the N--soliton solution as well.

The above results indicate that there exists a close relation between the KdV 
hierarchy and the SWW equation with $\alpha=2 \beta$. The root of this 
relation becomes more explicit in the AKNS scheme.$^{1)}$ Indeed, as a 
by--product of this scheme, it has been shown that the family of 
equations
\be
q_t + \hat{C}(4L) q_x = 0 \, ,
\label{evo}
\ee
where
\be
L = - \frac{1}{4} \frac{\partial^2}{\partial x^2} - q + \frac{1}{2} q_x 
\int_{x}^{\infty} dy \, ,
\ee
and $\hat{C}(k^2)=\omega(k)/k$, with $\omega(k)$ a given dispersion relation, 
is integrable, having all of them the same associated eigenvalue problem. In 
the specific case of $\hat{C}(k^2)=1/(1+k^2)$, Eq.(\ref{evo}) is just the SWW 
equation with $\alpha=2 \beta$. On the other hand, if we expand 
$\hat{C}(k^2)$ according to
\be
\hat{C}(k^2) = 1 - k^2 + k^4 - k^6 + \cdots \, ,
\ee
we get
\be
q_t + q_x - 4 L (q_x) + 16 L^2 (q_x) - \cdots = 0  \; .
\ee
When passing to the multiple--time slow variables $(q_0,\xi,\tau_{2n+1})$, 
with $q_0=\epsilon^{-2} q$, one can immediately see the emergence of the 
KdV hierarchy as
$$
q_{0 \tau_{2n+1}} = (4 L)^n q_{0 \xi}
$$
does generate the hierarchy.$^{13)}$ It should be mentioned that the above 
result is independent of any specific solution to the KdV hierarchy. 

\section{Final Remarks}
 
We have in this paper suceeded in writing the solitary--wave solution of the 
SWW equation (\ref{sww}), valid for any value of the parameters $\alpha$ and 
$\beta$, as a solitary--wave satisfying simultaneously, in the slow variables, 
all equations of the KdV hierarchy. This is actually a result valid for any 
system presenting an exact solitary--wave solution, that is, for 
systems whose solitary--wave initial condition does not radiate. It does not 
matter, in this case, whether the system is integrable or not.

We have then shown that, when  $v_0$ is assumed to be not a 
solitary--wave but a two or more soliton solution of the KdV hierarchy, the 
right--hand side of Eq.(\ref{v2b}), as well as all other corresponding higher 
order equations, does not vanish anymore, which in principle would mean 
that the perturbative series does not truncate. However, in the particular case 
of $\alpha = 2 \beta$, all those equations become homogeneous again, 
making then possible the truncation of the perturbative series through the 
choice of the trivial solutions
\be
v_n = 0 \quad ; \quad n = 2, 3, \dots \,\, .
\ee
The return to the laboratory coordinates leads $v_0$ to the two--or--more 
soliton solution of the SWW equation with $\alpha=2 \beta$, which ultimately 
means that the $N$--soliton solution of this SWW equation is nothing but a
$N$--soliton solution satisfying simultaneously all equations of the KdV 
hierarchy.  We have in this way shown that the dynamics underlying such 
solutions can be given in terms of that governing the N--soliton solutions of 
the whole KdV hierarchy, a fact that can also be seen to be true 
through the AKNS scheme. 

It is important to notice the difference between the cases $N=1$ and $N \geq 
2$. For $N=1$, the dynamics are equivalent for any $\alpha$ and $\beta$. 
However, for $N \geq 2$, the above equivalence exists only for $\alpha=2 
\beta$. The reason for this difference is the presence of soliton interactions in 
the case $N \geq 2$. As is well known, the SWW equation is in general 
non--integrable, which means that soliton interaction is inelastic. And it is 
exactly this inelastic character of the soliton interaction that will 
generate secularities other than those originated from the linear terms of the 
higher--order equations. What happens in the multiple scale method is that 
these new secularities, originated from the non--integrability of the original 
system, can not eliminated by the integrable equations of the KdV 
hierarchy. In this sense, the case with $\alpha=2 \beta$ is indeed peculiar, and 
can not be extended to the general case with any $\alpha$ and $\beta$. In 
other words, being the original system non--integrable, it can not be 
described by the integrable system formed by the equations of the KdV 
hierarchy.
 
\vspace{1 cm}
\section*{Acknowledgements}

The authors would like to thank J. L\'eon for useful discussions, and Y. 
Kodama and A. V. Mikhailov for sending the paper of Ref.[11] to them. This 
work was partially supported by CNPq (Brazil) and CNRS (France).  

\section*{References}
\noindent
1) M. J. Ablowitz, D. J. Kaup, A. C. Newell and H. Segur, Stud. Appl. Math. {\bf 
53}, 249 (1974). \\
\noindent
2) R. Hirota and J. Satsuma, J. Phys. Soc. Jpn. {\bf 40}, 611 (1976). \\
\noindent
3) P. A. Clarkson and E. L. Mansfield, Nonlinearity {\bf 7}, 975 (1994). \\
\noindent
4) A. Espinosa and J. Fujioka, J. Phys. Soc. Jpn. {\bf 63}, 1289 (1994). \\
\noindent
5) R. A. Kraenkel, M. A. Manna and J. G. Pereira, J. Math. Phys. {\bf 36}, 307 
(1995). \\
\noindent
6) R. A. Kraenkel, M. A. Manna, J. C. Montero and J. G. Pereira, J. Math. Phys. 
{\bf 36}, 6822 (1995). \\
\noindent
7) T. Taniuti, Suppl. Prog. Theor. Phys. {\bf 55}, 1 (1974). \\
\noindent
8) We thank R. Willox for useful comments on this point. \\
\noindent
9) R. A. Kraenkel, M. A. Manna, J. C. Montero and J. G. Pereira, Proceedings of 
the 
Workshop {\it Nonlinear Physics: Theory and Experiment}, Gallipoli, Lecce, 
1995 
(World Scientific, Singapore, 1995) (patt-sol/9509003). \\
\noindent
10) Y. Kodama and T. Taniuti, J. Phys. Soc. Jpn. {\bf 45}, 298 (1978). \\
\noindent
11) Y. Kodama and A. V. Mikhailov, {\it Obstacles to Asymptotic Integrability}, 
in {\it Algebraic
Aspects of Integrable Systems: in Memory of Irene Dorfman}, ed. by A. S. 
Fokas and I. Gel'fand 
(Birkhauser, 1996). \\
\noindent
12) R. A. Kraenkel, M. A. Manna, V. Merle, J. C. Montero and J. G. Pereira, 
Phys. Rev. E {\bf 54}, 2976 (1996). \\
\noindent 
13) M. J. Ablowitz and P. A. Clarkson, {\it Solitons, Nonlinear Evolution 
Equations and Inverse Scattering} 
(Cambridge University Press, Cambridge, 1992).

\end{document}